\documentclass[english,journal,12pt,onecolumn,draftclsnofoot]{IEEEtran}
\usepackage[T1]{fontenc}
\usepackage[latin9]{inputenc}
\usepackage{color}
\usepackage{amsmath}
\usepackage{amssymb}
\usepackage{graphicx}

\makeatletter

\providecommand{\tabularnewline}{\\}

\usepackage{flushend}

\makeatother

\usepackage{babel}
\begin{document}
\title{Low-Cost Uplink Sparse Code Multiple Access for Spatial Modulation}
\author{Ibrahim Al-Nahhal, \textit{Student Member, IEEE},\textit{ }Octavia
A. Dobre, \textit{Senior Member, IEEE}, Ertugrul Basar, \textit{Senior
Member, IEEE}, and Salama Ikki, \textit{Senior} \textit{Member, IEEE}}
\maketitle
\begin{abstract}
Spatial modulation (SM)-sparse code multiple access (SCMA) systems
provide high spectral efficiency (SE) at the expense of using a high
number of transmit antennas. To overcome this drawback, this letter
proposes a novel SM-SCMA system operating in uplink transmission,
referred to as rotational generalized SM-SCMA (RGSM-SCMA). For the
proposed system, the following are introduced: a) transmitter design
and its formulation, b) maximum likelihood and maximum a posteriori
probability decoders, and c) practical low-complexity message passing
algorithm and its complexity analysis. Simulation results and complexity
analysis show that the proposed RGSM-SCMA system delivers the same
SE with significant savings in the number of transmit antennas\textcolor{blue}{,}
at the expense of close bit error rate and a negligible increase in
the decoding complexity, when compared with SM-SCMA.
\end{abstract}


\section{Introduction}

\def\figurename{Fig.}
\def\tablename{TABLE}

\IEEEPARstart{S}{parse} code multiple access (SCMA) is a promising
non-orthogonal multiple access (NOMA) approach for 5G wireless networks
\cite{Mohammadkarimi_Octavia}-\cite{NOMA_Survey_2017} {\small{}that}
has been introduced in \cite{Nikopour_SCMA_2013}. SCMA assigns unique
multi-carrier sparse codes to each user to access the medium \cite{codebook_design_2014}.
The sparsity property of codes

\vspace{3mm}

{\footnotesize{}O. A. Dobre and I. Al-Nahhal are with the Faculty
of Engineering and Applied Science, Memorial University, St. John\textquoteright s,
NL, Canada (e-mail: \{odobre, ioalnahhal\}@mun.ca).}{\footnotesize\par}

{\footnotesize{}E. Basar is with the CoreLab, Department of Electrical
and Electronics Engineering, Ko\c{c} University, Istanbul, Turkey
(e-mail: ebasar@ku.edu.tr).}{\footnotesize\par}

{\footnotesize{}S. Ikki is with the Department of Electrical Engineering,
Lakehead University, Thunder Bay, ON, Canada (e-mail: sikki@lakeheadu.ca).}{\footnotesize\par}

\noindent enables the application of the message passing algorithm
(MPA) at the receiver, to provide near maximum likelihood (ML) bit
error rate (BER) performance with lower decoding complexity \cite{MPA_2015}.
The number of interfered users for each sub-carrier is also reduced,
allowing more users to be overloaded\textcolor{blue}{,} hence increasing
the spectral efficiency (SE) of the system.

Spatial modulation (SM) is another promising technique that provides
high SE with low-complexity signal detection \cite{Raed_SM_2008,SM_5G_Ertugrul_2016}.
It increases the SE by assigning part of the input data stream, named
spatial symbol, to activate an antenna to transmit the modulation
symbol. In \cite{GSM_Younis_2010}, generalized SM (GSM) is proposed
to overcome the limitation of the high number of transmit antennas
required in the SM system.

Recently, for further SE improvement, SM and NOMA have been jointly
considered \cite{SM_NOMA_Power_2018}-\cite{SM_SCMA_ltter_2018}.
Power-domain NOMA, low-density signature, and SCMA have been explored
for SM in \cite{SM_NOMA_Power_2018}, \cite{SM_SCMA_LDS_2018}, and
\cite{SM_SCMA_ltter_2018}, respectively. Such systems require an
integer power of two transmit antennas to deliver spatial symbols,
which comes to be infeasible for higher rate transmission.

This letter proposes a novel uplink SM system, referred to as rotational
GSM (RGSM)-SCMA, which overcomes the previously mentioned drawback
of the existing SM-NOMA systems. The models of the proposed RGSM-SCMA
transmitter and receiver for the uplink scenario are introduced. ML
and maximum \textit{a posteriori} probability (MAP) decoders are provided
as theoretical receivers. Additionally, the iterative MPA decoder
is presented and analyzed to provide a practical low-complexity detection.
The proposed RGSM-SCMA system enjoys a high SE transmission as the
SM-SCMA system with a significant reduction in the number of transmit
antennas, which leads to saving resources that can then be used for
channel estimation. It is shown that the MPA decoder for the proposed
RGSM-SCMA system attains a close BER performance to the MPA of the
SM-SCMA system, with nearly the same complexity.

\section{Related Work and Motivation}

In a single-user SM system, the input data stream is transmitted as
a combination of spatial symbols and modulated symbols. At the receiver
side, the decoder estimates both the spatial and modulated symbols
by performing an exhaustive search or by using one of the low-complexity
decoding algorithms, such as those in \cite{QSM_Letter}-\cite{SM_JSAC}.
For instance, assume that an SM system is equipped with four transmit
antennas (i.e., four spatial symbols) and two modulated symbols (i.e.,
binary phase shift keying). Thus, this system can deliver 3-bits at
a time; 2-bits spatial symbol (i.e., $\text{log}_{2}(\text{number of transmit antennas}$)-bits
which corresponding to each of the four antennas) and 1-bit modulated
symbol. As seen from this example, the number of transmit antenna
must be a power of two, which increases exponentially as the SE increases.

For a multi-user SM system, the SCMA technique is used to organize
the accessing of the users to the medium. This system is known as
SM-SCMA. The SM-SCMA system enjoys a high SE with good BER performance,
which is suitable for the future generations of the wireless networks.
However, the use of high number of transmit antennas is still required.

In this paper, we propose a solution to this problem by activating
more than one antenna at a time to deliver the same SE of SM-SCMA
with a much lower number of transmit antennas. In addition, rotational
angles are used to provide a close BER performance to the SM-SCMA.

\section{RGSM-SCMA System Model}

\begin{figure*}[t]
\begin{centering}
\includegraphics[scale=0.45]{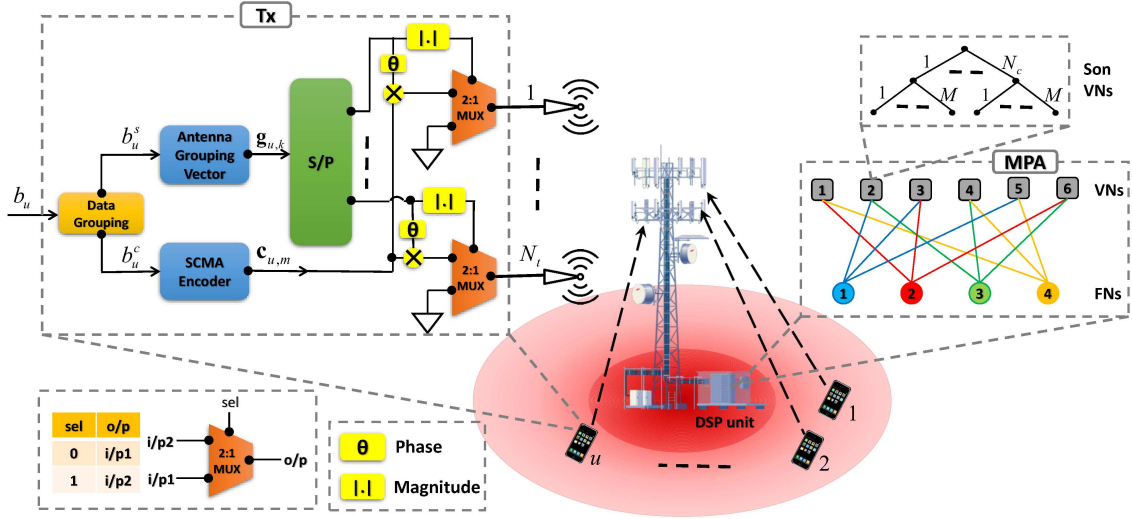}
\par\end{centering}
\caption{\label{fig: Block_Diagram}{\small{}Uplink RGSM-SCMA block diagram
for the $u$-th user}.}
\end{figure*}

In this section, the RGSM-SCMA system is introduced. Assume that $R$
orthogonal resource elements (OREs), e.g., subcarriers, are overloaded
with $U$ users (i.e., $U>R$); each user has a unique sparse codebook,
$\mathbf{C}_{u}\in\mathbb{C}^{R\times M}$, $u=1,\ldots,U$, which
contains $M$ codewords, $\mathbf{c}_{u,m}\in\mathbb{C}^{R\times1}$,
$m=1,\ldots,M$. $\mathbf{c}_{u,m}$ has $d_{v}$ non-zero codeword
elements in the same positions for each codebook, and vary from one
codebook to another. The number of the overlapped users per ORE, $d_{f}$,
is fixed for $\forall R$. The SE for the $u$-th user is $\eta_{u}=\eta_{u}^{s}+\eta_{u}^{c}$
bit per channel use (bpcu), where $\eta_{u}^{s}$ and $\eta_{u}^{c}$
denote the spatial and code spectral efficiencies, respectively. 

Consider that $N_{t}$ and $M$ are the number of transmit antennas
used to deliver $\eta_{u}^{s}$ bpcu and the number of codewords used
to deliver $\eta_{u}^{c}$ bpcu, respectively, for each user. It is
assumed that the system parameters for all users are the same (i.e.,
same $N_{t}$, SEs and $M$). In the RGSM-SCMA system, $\eta_{u}^{s}=\log_{2}(N_{c})$
bpcu, where $N_{c}=2^{n_{c}}$ antenna combinations, $n_{c}=${\footnotesize{}$\left\lfloor \log_{2}\left(\begin{array}{c}
N_{t}\\
N_{a}
\end{array}\right)\right\rfloor $}, with $\left\lfloor \cdot\right\rfloor $ as the floor operation,
and $N_{a}<N_{t}$ is the number of active antennas at a time, while
$\eta_{u}^{c}=\log_{2}(M)$ bpcu.

Fig. \ref{fig: Block_Diagram} shows the uplink scenario of the RGSM-SCMA
system for the $u$-th user; the input $\mathbf{b}_{u}\in\mathbb{B}^{\eta_{u}}$
bits for the $u$-th user is divided into two parts: the first $\mathbf{b}_{u}^{s}\in\mathbb{B}^{\eta_{u}^{s}}$
bits represent the spatial symbol, while the last $\mathbf{b}_{u}^{c}\in\mathbb{B}^{\eta_{u}^{c}}$
bits represent the code symbol. The SCMA encoder block maps the $\mathbf{b}_{u}^{c}$
bits to its corresponding codeword $\mathbf{c}_{u,m}$ and delivers
it to the input of all transmit antenna multiplexers. The antenna
grouping vector block chooses the antenna grouping vector, $\mathbf{g}_{u,k}\in\mathbb{C}^{1\times N_{t}}$,
$k\in\{1,\ldots,N_{c}\}$, according to the value of $\mathbf{b}_{u}^{s}$
from a predetermined lookup table (Table \ref{tab:SCMA-GSM-antenna-grouping}
is an example lookup table with $N_{t}=5$, $N_{a}=2$, $N_{c}=8$,
and $\mathbf{b}_{u}^{s}=3$ bpcu). It should be noted that $\mathbf{g}_{u,k}$
has $N_{a}$ non-zero elements that correspond to the $N_{a}$ active
antennas. The serial-to-parallel (S/P) block distributes the zero
and non-zero elements of $\mathbf{g}_{u,k}$ at the same time, to
the next stage. The magnitude of $\mathbf{g}_{u,k}$, $\left|\mathbf{g}_{u,k}\right|\in\mathbb{R}^{1\times N_{t}}$,
is applied to the multiplexers' selector pins, which allow the antennas
corresponding to the non-zero elements to transmit $\mathbf{c}_{u,m}$
rotated with the associated rotation angle of $\mathbf{g}_{u,k}$.
To ensure maximum distance between the successive angles, they should
be equally spaced for each antenna. Thus, the rotation angles of the
$n_{t}$-th antenna are

\begin{table}
\caption{\label{tab:SCMA-GSM-antenna-grouping}{\small{}The RGSM-SCMA antenna
grouping vector lookup table for $N_{t}=5$, $N_{a}=2$, $N_{c}=8$,
and $\eta_{u}^{s}=3$ bpcu}. }

\centering{}%
\begin{tabular}{c|c|c}
\hline 
$\mathbf{b}_{u}^{s}$ & $k$ & $\mathbf{g}_{u.k}$\tabularnewline
\hline 
\hline 
$000$ & $1$ & $[\begin{array}{ccccc}
1\,\,\,\hspace{0.5cm} & 0\,\,\,\hspace{0.5cm} & \,\,\,1\hspace{0.5cm} & \hspace{0.5cm}0\,\,\,\,\,\,\,\hspace{0.5cm} & \hspace{0.5cm}0\end{array}]$\tabularnewline
$001$ & $2$ & $[\begin{array}{ccccc}
\text{e}^{-j\frac{2\pi}{3}} & 0\,\,\,\hspace{0.5cm} & \,\,\,0\hspace{0.5cm} & \hspace{0.5cm}1\,\,\,\,\,\,\,\hspace{0.5cm} & \hspace{0.5cm}0\end{array}]$\tabularnewline
$010$ & $3$ & $[\begin{array}{ccccc}
\text{e}^{-j\frac{4\pi}{3}} & 0\,\,\,\hspace{0.5cm} & \,\,\,0\hspace{0.5cm} & \hspace{0.5cm}0\,\,\,\,\,\,\,\hspace{0.5cm} & \hspace{0.5cm}1\end{array}]$\tabularnewline
$011$ & $4$ & $[\begin{array}{ccccc}
0\hspace{0.5cm}\, & \,\,\,1\,\,\hspace{0.5cm} & \,\,\,\,\text{e}^{-j\frac{\pi}{2}} & \,\,\,\,\,\,\,\,\,0\,\,\,\,\,\,\,\hspace{0.5cm} & \hspace{0.5cm}0\end{array}]$\tabularnewline
$100$ & $5$ & $[\begin{array}{ccccc}
0\hspace{0.5cm}\, & \,\,\,\text{e}^{-j\frac{2\pi}{3}} & \,\,0\,\hspace{0.5cm}\hspace{0.4cm} & \,\text{e}^{-j\frac{2\pi}{3}}\,\,\,\, & \hspace{0.5cm}0\end{array}]$\tabularnewline
$101$ & $6$ & $[\begin{array}{ccccc}
0\hspace{0.5cm} & \,\,\,\,\text{e}^{-j\frac{4\pi}{3}}\, & \,0\,\,\,\,\,\,\,\,\,\,\,\,\,\,\,\,\,\,\, & \,0\,\,\,\,\hspace{0.5cm} & \text{e}^{-j\frac{2\pi}{3}}\end{array}]$\tabularnewline
$110$ & $7$ & $[\begin{array}{ccccc}
0\hspace{0.5cm}\, & \,\,\,0\hspace{0.5cm} & \,\,\,\,\,\,\text{e}^{-j\pi}\hspace{0.5cm} & \text{e}^{-j\frac{4\pi}{3}}\,\,\,\, & \hspace{0.5cm}0\end{array}]$\tabularnewline
$111$ & $8$ & $[\begin{array}{ccccc}
0\,\,\,\,\,\,\,\,\, & \,\,\,\,\,0\,\,\,\,\,\, & \hspace{0.5cm}\text{e}^{-j\frac{3\pi}{2}} & \,\,\,\,\,\,0\,\,\,\,\hspace{0.5cm} & \text{e}^{-j\frac{4\pi}{3}}\end{array}]$\tabularnewline
\hline 
\end{tabular}
\end{table}

\begin{equation}
\theta_{d,n_{t}}\hspace{-0.1cm}=\frac{-2\left(d-1\right)\pi}{a_{n_{t}}},\,\,\,n_{t}=1,\ldots,N_{t},\,\,d=1,\ldots,a_{n_{t}},\label{eq:Rotational_Angles}
\end{equation}

\noindent where $\theta_{d,n_{t}}$ denotes the rotation angle of
the $d$-th occurrence of the $n_{t}$-th antenna, and $a_{n_{t}}$
is the number of times that the $n_{t}$-th antenna is activated for
$\forall k$. For instance, in Table \ref{tab:SCMA-GSM-antenna-grouping},
the third antenna (i.e., $n_{t}=3$) is activated 4 times (i.e., $a_{3}=4$)
at $k=1,\,4,\,7$ and $8$. Then, $\theta_{1,3}$, $\theta_{2,3}$,
$\theta_{3,3}$ and $\theta_{4,3}$ equal to $0$, $\frac{-\pi}{2}$,
$-\pi$ and $\frac{-3\pi}{2}$, respectively.

At the receiver side, the noisy received signal for each ORE at the
$n$-th receive antenna, $y_{n}^{r}$, is

\begin{equation}
y_{n}^{r}=\sum_{u\in\varLambda_{r}}\left(\mathbf{h}_{u,n}^{r}\mathbf{g}_{u,k}^{\text{T}}c_{u,m}^{r}\right)+n_{n}^{r},\label{eq: y_n_r}
\end{equation}

\noindent where $\mathbf{h}_{u,n}^{r}\in\mathbb{C}^{1\times N_{t}}$
denotes the Rayleigh fading channel between the $N_{t}$ transmit
antennas and $n$-th receive antenna of the $u$-th user for the $r$-th
ORE, $c_{u,m}^{r}$ represents the $r$-th element of the $m$-th
codeword for user $u$, $n_{n}^{r}\sim\mathcal{N}\left(0,\sigma^{2}\right)$
is the Gaussian noise at the $r$-th ORE of the $n$-th receive antenna
with zero-mean and a variance of $\sigma^{2}$, and $\varLambda_{r}$
is the set of indices of the users that share the $r$-th ORE. The
received signals vector, $\mathbf{y}_{n}$, for all OREs at the $n$-th
receive antenna is

\begin{equation}
\mathbf{y}_{n}=\sum_{u=1}^{U}\left(\mathrm{diag}\left(\mathbf{h}_{u,n}\mathbf{g}_{u,k}^{\text{T}}\right)\mathbf{c}_{u,m}\right)+\mathbf{n}_{n},\label{eq: y_n}
\end{equation}

\noindent where $\mathbf{y}_{n}\in\mathbb{C}^{R\times1}=\left[y_{n}^{1},\ldots,y_{n}^{R}\right]^{\text{T}}$,
$\mathbf{h}_{u,n}\in\mathbb{C}^{R\times N_{t}}=\left[\mathbf{h}_{u,n}^{1\,T},\ldots,\mathbf{h}_{u,n}^{R\,T}\right]^{\text{T}}$,
$\mathbf{n}_{n}\in\mathbb{C}^{R\times1}=\left[n_{n}^{1}\ldots n_{n}^{R}\right]^{\text{T}}$,
and $\mathrm{diag}\left(\mathbf{h}_{u,n}\mathbf{g}_{u,k}^{\text{T}}\right)\in\mathbb{C}^{R\times R}$
is a diagonal matrix whose $r$-th diagonal element is $\mathbf{h}_{u,n}^{r}\mathbf{g}_{u,k}^{\text{T}}$.

\section{RGSM-SCMA Signal Detection}

In this section, the formulation of three decoders for the proposed
RGSM-SCMA system is deduced, which are ML, MAP, and MPA decoders.

\subsection{ML Decoder}

The ML decoder performs an exhaustive search for all $\left(N_{c}M\right)^{U}$
possibilities to provide the optimum BER performance. The ML solution
for $N_{r}$ receive antennas is 

$\vphantom{}$

$\left\{ \hat{\mathbf{C}},\hat{\mathbf{G}}\right\} =\underset{\begin{array}{c}
j\in N_{c}^{U}\\
l\in M^{U}
\end{array}}{\text{arg}\,\text{min}}$

\begin{equation}
\left\{ \sum_{n=1}^{N_{r}}\left\Vert \mathbf{y}_{n}-\sum_{u=1}^{U}\left(\text{diag}\left(\mathbf{h}_{u,n}\mathbf{g}_{u,k(j)}^{\text{T}}\right)\mathbf{c}_{u,m(l)}\right)\right\Vert ^{2}\right\} .\label{eq:ML}
\end{equation}

\noindent Here, $\hat{\mathbf{C}}\in\mathbb{C}^{R\times U}=\left[\hat{\mathbf{c}}_{1,m}\ldots\hat{\mathbf{c}}_{U,m}\right]$
denotes the estimated transmitted codewords for all users, with $\hat{\mathbf{c}}_{u,m}$
as the estimated transmitted codeword for the $u$-th user, and $m(l)$
represents the value of $m\in\left\{ 1,\ldots,M\right\} $ at the
$l$-th antenna combination. $\hat{\mathbf{G}}\in\mathbb{C}^{U\times N_{t}}=\left[\hat{\mathbf{g}}_{1,k}\ldots\hat{\mathbf{g}}_{U,k}\right]$
denotes the estimated grouping vectors for all users, with $\hat{\mathbf{g}}_{u,k}$
as the estimated grouping vector for the $u$-th user, and $k(j)$
represents the value of $k\in\left\{ 1,\ldots,N_{c}\right\} $ at
the $j$-th antenna combination.

\subsection{MAP Decoder}

Unlike the ML decoder, the MAP decoder estimates the pair of transmitted
codeword and grouping vector, $\{\hat{\mathbf{c}}_{u,m},\hat{\mathbf{g}}_{u,k}\}$,
for each user one-by-one by maximizing \emph{a posteriori} probability
of this pair given the received signal as

\[
\left\{ \hat{\mathbf{c}}_{u,m},\hat{\mathbf{g}}_{u,k}\right\} =\underset{\begin{array}{c}
\hat{\mathbf{c}}_{u,m}\in\mathbf{C}_{u}\\
\hat{\mathbf{g}}_{u,k}\in\mathbf{G}_{u}
\end{array}}{\text{arg}\,\text{max}}\left\{ P\left(\{\hat{\mathbf{c}}_{u,m},\hat{\mathbf{g}}_{u,k}\}|\mathbf{y}_{n}\right)\right\} 
\]

\begin{multline*}
=\underset{\begin{array}{c}
\hat{\mathbf{c}}_{u,m}\in\mathbf{C}_{u}\\
\hat{\mathbf{g}}_{u,k}\in\mathbf{G}_{u}
\end{array}}{\text{arg}\,\text{max}}\left\{ P\left(\{\hat{\mathbf{c}}_{u,m},\hat{\mathbf{g}}_{u,k}\}\right)P\left(\mathbf{y}_{n}|\{\hat{\mathbf{c}}_{u,m},\hat{\mathbf{g}}_{u,k}\}\right)\right\} 
\end{multline*}

\begin{multline*}
=\hspace{-0.3cm}\underset{\begin{array}{c}
\hat{\mathbf{c}}_{u,m}\in\mathbf{C}_{u}\\
\hat{\mathbf{g}}_{u,k}\in\mathbf{G}_{u}
\end{array}}{\text{arg}\,\text{max}}\hspace{-0.4cm}\{\sum_{\begin{array}{c}
\left.\mathbf{\exists}\right\backslash \{\hat{\mathbf{c}}_{u,m},\hat{\mathbf{g}}_{u,k}\}\\
\,\in\left.\mathbf{\Xi}\right\backslash \{\mathbf{C}_{u},\mathbf{G}_{u}\}
\end{array}}\hspace{-1cm}P\left(\{\hat{\mathbf{c}}_{u,m},\hat{\mathbf{g}}_{u,k}\},\left.\mathbf{\exists}\right\backslash \{\hat{\mathbf{c}}_{u,m},\hat{\mathbf{g}}_{u,k}\}\right)
\end{multline*}

\begin{equation}
\hspace{2cm}\times P\left(\mathbf{y}_{n}|\{\hat{\mathbf{c}}_{u,m},\hat{\mathbf{g}}_{u,k}\},\left.\mathbf{\exists}\right\backslash \{\hat{\mathbf{c}}_{u,m},\hat{\mathbf{g}}_{u,k}\}\right)\},\label{eq: MAP_Bayes}
\end{equation}

\noindent where $\mathbf{\exists}=\left[\{\hat{\mathbf{c}}_{1,m},\hat{\mathbf{g}}_{1,k}\},\ldots,\{\hat{\mathbf{c}}_{U,m},\hat{\mathbf{g}}_{U,k}\}\right]$
represents one possible combination of the transmitted set of codewords
and grouping vectors for all $U$ users, $\mathbf{\Xi}$ is the set
containing all $\left(N_{c}M\right)^{U}$ possibilities of $\mathbf{\exists}$,
$\left.\mathbf{\exists}\right\backslash \{\hat{\mathbf{c}}_{u,m},\hat{\mathbf{g}}_{u,k}\}$
denotes $\mathbf{\exists}$ except the set $\{\hat{\mathbf{c}}_{u,m},\hat{\mathbf{g}}_{u,k}\}$,
and $\left.\mathbf{\Xi}\right\backslash \{\mathbf{C}_{u},\mathbf{G}_{u}\}$
denotes $\mathbf{\Xi}$ except $\{\mathbf{C}_{u},\mathbf{G}_{u}\}$.
Note that the total probability theorem is applied to obtain the last
line of (\ref{eq: MAP_Bayes}). Since all elements of $\mathbf{y}_{n}$
are independent for all OREs and $N_{r}$ receive antennas, the conditional
probability in the last term of (\ref{eq: MAP_Bayes}) becomes

\begin{equation}
P\left(\mathbf{y}_{n}|\{\hat{\mathbf{c}}_{u,m},\hat{\mathbf{g}}_{u,k}\},\left.\mathbf{\exists}\right\backslash \{\hat{\mathbf{c}}_{u,m},\hat{\mathbf{g}}_{u,k}\}\right)=\hspace{-0.1cm}\prod_{n=1}^{N_{r}}\prod_{r\in\Omega_{u}}P\left(\mathbf{y}_{n}|\mathbf{\exists}(r)\right),\label{eq: MAP_independent}
\end{equation}

\noindent where $\Omega_{u}$ is the set of indices of the $d_{v}$
non-zero OREs for the $u$-th user, $\mathbf{\exists}(r)$ denotes
one possible set of transmitted codewords and grouping vectors at
the $r$-th ORE for the $d_{f}$ users that share the $r$-th ORE,
and 

\begin{equation}
P\left(\mathbf{y}_{n}|\mathbf{\exists}(r)\right)\hspace{-0.1cm}=\hspace{-0.1cm}\frac{1}{\sqrt{2\pi}\sigma}\text{exp}\hspace{-0.1cm}\left(\hspace{-0.1cm}-\frac{\left|y_{n}^{r}-\hspace{-0.1cm}\sum_{u\in\varLambda_{r}}\hspace{-0.1cm}\left(\mathbf{h}_{u,n}^{r}\mathbf{g}_{u,k}^{\text{T}}c_{u,m}^{r}\right)\right|^{2}}{2\sigma^{2}}\right)\hspace{-0.1cm}.\label{eq:Condtional_Prob}
\end{equation}

\subsection{MPA Decoder}

\begin{table*}
\caption{\label{tab:Complexity}{\small{}The real operations of the MPA decoders
for the SM-SCMA and RGSM-SCMA systems}.}

\centering{}%
\begin{tabular}{c|c|c}
\cline{2-3} \cline{3-3} 
 & SM-SCMA & RGSM-SCMA\tabularnewline
\hline 
\hline 
Additions & $\begin{array}{c}
Rd_{f}\left(N_{t}M\right)^{d_{f}}\left(2N_{r}(2d_{f}+1)-1\right)\\
+TRd_{f}\left(\left(N_{t}M\right)^{d_{f}}-1\right)
\end{array}$ & $\begin{array}{c}
Rd_{f}\left(N_{c}M\right)^{d_{f}}\left(2N_{r}(2d_{f}+1)-1\right)\\
+TRd_{f}\left(\left(N_{c}M\right)^{d_{f}}-1\right)+2Ud_{v}N_{r}N_{c}\left(2N_{a}-1\right)
\end{array}$\tabularnewline
\hline 
Multiplications & $\begin{array}{c}
Rd_{f}\left(N_{t}M\right)^{d_{f}}\left(2N_{r}(2d_{f}+1)+Td_{f}+1\right)\\
+N_{t}M\left(d_{v}-1\right)\left(TRd_{f}+U\right)
\end{array}$ & $\begin{array}{c}
Rd_{f}\left(N_{c}M\right)^{d_{f}}\left(2N_{r}(2d_{f}+1)+Td_{f}+1\right)\\
+N_{c}M\left(d_{v}-1\right)\left(TRd_{f}+U\right)+4Ud_{v}N_{r}N_{c}N_{a}
\end{array}$\tabularnewline
\hline 
\end{tabular}
\end{table*}

The MPA decoder provides an approximation to the MAP detector using
the factor graph method, shown in Fig. \ref{fig: Block_Diagram}.
In the factor graph, the OREs and served users are represented as
function nodes (FNs) and variable nodes (VNs), respectively. For each
FN, all VNs that share this FN are connected. Note that each VN has
$N_{c}M$ son VNs. The idea of the MPA is to iteratively update the
probability of passing the messages from FNs to VNs and vice versa.
After $T$ iterations, the MPA stops and detects the message which
corresponds to the maximum joint probability. Note that the conventional
MPA of the proposed RGSM-SCMA is modified to jointly estimate the
antenna grouping vector and transmitted codeword.

To formulate the MPA, assume that $\mathcal{P}_{f_{r}\rightarrow v_{u}}^{(t)}(\{c_{u,m}^{r},\mathbf{g}_{u,k}\})$
and $\mathcal{P}_{v_{u}\rightarrow f_{r}}^{(t)}(\{c_{u,m}^{r},\mathbf{g}_{u,k}\})$
is the probability of passing the message $\{c_{u,m}^{r},\mathbf{g}_{u,k}\}$
from the $r$-th FN to the $u$-th VN and from the $u$-th VN to the
$r$-th FN, respectively, at the $t$-th iteration, $t=1,\ldots,T$.
First, all messages sent from VNs to FNs are assumed equiprobable
at the first iteration; i.e.,

\begin{equation}
\mathcal{P}_{v_{u}\rightarrow f_{r}}^{(0)}\left(\{c_{u,m}^{r},\mathbf{g}_{u,k}\}\right)=\frac{1}{N_{c}M},\,\,\,\,\,\forall u,\,\,\forall r,\,\,\forall m,\,\,\forall k.\label{eq:Equal_prob}
\end{equation}

\noindent Now, $\mathcal{P}_{f_{r}\rightarrow v_{u}}^{(t+1)}(\{c_{u,m}^{r},\mathbf{g}_{u,k}\})$
can be written as

\[
\mathcal{P}_{f_{r}\rightarrow v_{u}}^{(t+1)}\left(\{c_{u,m}^{r},\mathbf{g}_{u,k}\}\right)=\,\hspace{5cm}
\]

\[
\sum_{\mathbf{\exists}(i),i\in\left.\varLambda_{r}\right\backslash u}\left\{ \prod_{n=1}^{N_{r}}\left(P\left(\mathbf{y}_{n}|\mathbf{\exists}(i),\mathbf{\exists}(u)=\{c_{u,m}^{r},\mathbf{g}_{u,k}\}\right)\right)\right.
\]

\begin{equation}
\left.\times\prod_{i\in\left.\varLambda_{r}\right\backslash u}\mathcal{P}_{v_{i}\rightarrow f_{r}}^{(t)}\left(\mathbf{\exists}(i)\right)\right\} ,\,\,\,\,\,\forall m,\,\,\forall k,\,\,\forall r,\,\,u\in\varLambda_{r},\label{eq:FN_to_VN}
\end{equation}

\noindent where $\left.\varLambda_{r}\right\backslash u$ denotes
$\varLambda_{r}$ except the $u$-th user, and $P\left(\mathbf{y}_{n}|\mathbf{\exists}(i),\mathbf{\exists}(u)=\{c_{u,m}^{r},\mathbf{g}_{u,k}\}\right)$
is given in (\ref{eq:Condtional_Prob}). Then, the probability of
passing the messages from VNs to FNs is updated as

\[
\mathcal{P}_{v_{u}\rightarrow f_{r}}^{(t+1)}\left(\{c_{u,m}^{r},\mathbf{g}_{u,k}\}\right)=\gamma_{u,r}^{(t+1)}\hspace{-0.2cm}\prod_{j\in\left.\Omega_{u}\right\backslash r}\hspace{-0.2cm}\mathcal{P}_{f_{j}\rightarrow v_{u}}^{(t+1)}\left(\{c_{u,m}^{r},\mathbf{g}_{u,k}\}\right),
\]

\begin{equation}
\hspace{4cm}\forall m,\,\,\forall k,\,\,\forall u,\,\,r\in\Omega_{u},\label{eq:VN_to_FN}
\end{equation}

\noindent where $\left.\Omega_{u}\right\backslash r$ denotes $\Omega_{u}$
except the $r$-th ORE, and $\gamma_{u,r}^{(t+1)}$ is the normalization
factor, which is given by

\begin{equation}
\gamma_{u,r}^{(t+1)}=\left(\sum_{m=1}^{M}\sum_{k=1}^{N_{c}}\mathcal{P}_{v_{u}\rightarrow f_{r}}^{(t)}\left(\{c_{u,m}^{r},\mathbf{g}_{u,k}\}\right)\right)^{-1}.
\end{equation}

\noindent After $T$ iterations, the estimated transmitted codeword
and grouping vector are obtained as

\begin{equation}
\left\{ \hat{\mathbf{c}}_{u,m},\hat{\mathbf{g}}_{u,k}\right\} \hspace{-0.1cm}=\hspace{-0.4cm}\hspace{-0.2cm}\underset{\begin{array}{c}
m=1,\ldots,M\\
k=1,\ldots,N_{c}
\end{array}}{\text{arg}\,\text{max}}\hspace{-0.2cm}\prod_{j\in\Omega_{u}}\mathcal{P}_{f_{j}\rightarrow v_{u}}^{(T)}\left(\{\mathbf{c}_{u,m},\mathbf{g}_{u,k}\}\right),\,\,\,\forall u.\label{eq: Final_MPA}
\end{equation}

\subsection{MPA Complexity Analysis}

In this subsection, the computational complexity of the MPA decoders
for the SM-SCMA and RGSM-SCMA systems is deduced in terms of real
additions and multiplications. Table \ref{tab:Complexity} shows the
complexity summary, calculated based on (\ref{eq:Equal_prob})-(\ref{eq: Final_MPA})
and the factor graphs of both systems. At the same SE, it is shown
from Table \ref{tab:Complexity} that there is a negligible increase
in the number of real multiplications and additions of the RGSM-SCMA
by $4Ud_{v}N_{r}N_{c}N_{a}$ and $2Ud_{v}N_{r}N_{c}\left(2N_{a}-1\right)$,
respectively. This increase is a result of combining the channel entries
by the antenna grouping vectors before performing the MPA decoder,
and is independent of the number of iterations. Thus, at the same
SE, the decoding complexity of both systems is almost similar.

\section{Simulation Results}

\begin{figure}[t]
\begin{centering}
\includegraphics[scale=0.4]{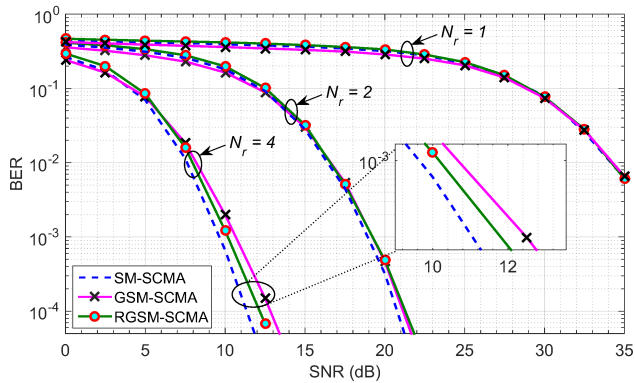}
\par\end{centering}
\caption{\label{fig: BER}{\small{}BER performance comparison}.}
\end{figure}

In this section, simulation is used to study the BER performance of
the proposed RGSM-SCMA system, additionally in comparison with the
SM-SCMA \cite{SM_SCMA_ltter_2018}. The effect of the rotation angles
in (\ref{eq:Rotational_Angles}) is shown, and we refer to the zero
rotation angles version of the RGSM-SCMA as GSM-SCMA. The MPA decoder
is considered for all systems. Furthermore, the Rayleigh fading channel
is assumed to be perfectly known at the receiver. The required number
of transmit antennas and decoding complexity comparisons between the
proposed RGSM-SCMA and SM-SCMA are also provided. The system parameters
for all systems are chosen as follows: $U=6$, $R=4$ and $M=4$.

Fig. \ref{fig: BER} shows the BER performance comparison for $\eta_{u}^{s}=3$
bpcu ($N_{t}=8$ and $5$ for SM-SCMA and RGSM-SCMA, respectively),
$N_{a}=2$ and $T=2$ in case of $N_{r}=1,\,2$, and $4$. As shown
in this figure, the BER performance is almost the same in the case
of $N_{r}=1$ and $2$. In the case of high signal-to-noise ratio
(SNR) for $N_{r}=4$, the proposed RGSM-SCMA provides a better BER
performance than the GSM-SCMA. Thus, using the rotation angles in
(\ref{eq:Rotational_Angles}) for the proposed RGSM-SCMA degrades
the BER performance by only $0.6$ dB instead of $1.2$ dB SNR as
in GSM-SCMA, when compared with SM-SCMA.

Fig. \ref{fig:Comp} shows the extra complexity (ExCo) of the RGSM-SCMA
over SM-SCMA mentioned in Table \ref{tab:Complexity} and given by:
ExCo = (RGSM operations - SM operations) / SM operations, where operations
can be either additions or multiplications. It can be seen from Fig.
\ref{fig:Comp} that the ExCo is negligible (less than $0.05\%$)
for both additions and multiplications, and decreases when $T$ or
$\eta_{u}^{s}$ increase. 

The RGSM-SCMA system provides significant savings in the number of
transmit antennas, $N_{t}$, required to deliver the same $\eta_{u}^{s}$
of the SM-SCMA, as shown in Fig. \ref{fig:Antenna}. Note that $N_{a}=\{2,2,3,3,4,4,4,5,5\}$
is used to achieve $\eta_{u}^{s}=2:10$ bpcu. The RGSM-SCMA saves
more in terms of $N_{t}$ when $\eta_{u}^{s}$ increases. For example,
to deliver $\eta_{u}^{s}=7$ bpcu, the SM-SCMA requires $128$ transmit
antennas, while the RGSM-SCMA requires only $10$ antennas.

Finally, the RGSM-SCMA provides a significant reduction in the number
of transmit antennas with almost the same decoding complexity and
a very slight deterioration in the BER performance to deliver the
same SE of the SM-SCMA.
\begin{center}
\begin{figure}
\begin{centering}
\includegraphics[scale=0.41]{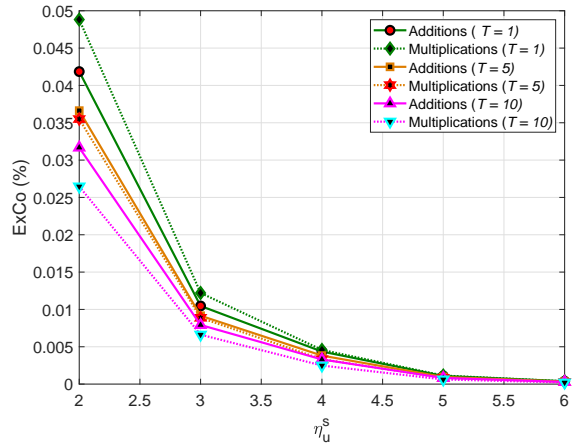}
\par\end{centering}
\caption{\label{fig:Comp}Extra complexity comparison between SM-SCMA and RGSM-SCMA.}
\end{figure}
\par\end{center}

\begin{center}
\begin{figure}
\begin{centering}
\includegraphics[scale=0.41]{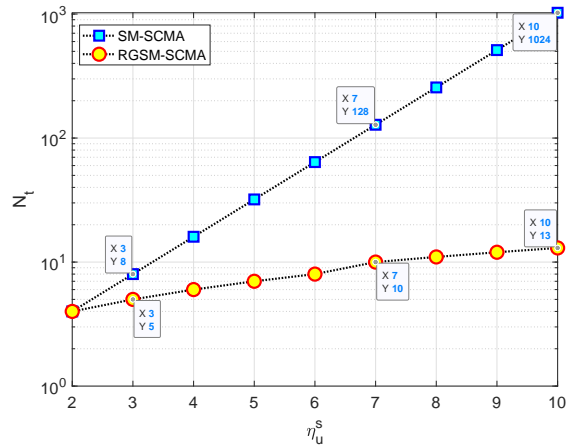}
\par\end{centering}
\caption{\label{fig:Antenna}$N_{t}$ comparison between SM-SCMA and RGSM-SCMA.}
\end{figure}
\par\end{center}

\section{Conclusion}

A low-cost SM-SCMA system has been proposed, which utilizes a reduced
number of transmit antennas, referred to as RGSM-SCMA. The transmitter
design, as well as the ML and MAP decoders have been introduced. Furthermore,
the low-complexity MPA decoder has been revised and analyzed for the
proposed RGSM-SCMA system. This delivers the same SE as SM-SCMA with
a much lower number of required antennas, at the expense of less than
$0.05\%$ increase in the decoding complexity and up to $0.6$ dB
SNR degradation in the BER performance.

\end{document}